# A Generalized Approach to Determination of Magnetic Shielding Factor for Physics Package of Rb Atomic Clock

S. S. Raghuwanshi and G.M.Saxena*

National Physical Laboratory

Dr K.S.Krishnan Road New Delhi-110012

* E mail: gmsaxena@nplindia.org

**Abstract**: In this paper we report generalized approach to calculate magnetic shielding factor (MSF) of multi-layer mu metal concentric cylindrical shields for arbitrary length to radius ratios and different values of magnetic permeability. We report in this paper the generalized results on the magnetic shielding factor of multi-layered magnetic shields used in Rb atomic clocks.

**Introduction:** In Rb atomic clock the Physics Package has a very critical role in deciding the frequency stability and the accuracy. In the Physics Package the hyperfine transitions in the ground state of Rb atoms take place and the necessary clock signal is produced. The hyperfine levels are affected by the magnetic field. It is necessary that the Rb atoms are shielded from the external magnetic fields and its fluctuations. In general multi layers of the magnetic shields are used for this purpose. In the case of $Rb^{87}$ atoms it has been shown[1] that a fractional frequency shift $\Delta f/f$ due to the magnetic field fluctuations $\Delta B$ in the applied d.c magnetic field B of the order of $2.0 \times 10^{-5}$ Tesla is given by the relation

$$\Delta f/f = 3.36 \times 10^{-4} \Delta B. \qquad (1)$$

It may be readily verified from this relation that for the fractional frequency stability of $10^{-13}$ a shielding factor of the order of $10^4$ is required. This shielding factor may be achieved by multi-layers of the high mu metal concentric cylindrical shields. The necessary computer simulations for obtaining generalized results on the shielding factor for various ratios of length to radius, magnetic permeability and demagnetization factor have been done and reported in the paper.

The magnetic shielding factor for a single layer mu metal cylindrical shield is defined as

$$S = \frac{\text{Increment in external field}}{\text{Increment in internal field}} \qquad (2)$$

For a single cylindrical shield with end caps the shielding factor is given by

$$S_t = \frac{1}{2}\mu t' \qquad (3a)$$

$$S_l = 2D\mu t'(1 + R/L)^{-1} \qquad (3b)$$

Here the subscripts t and l stand for transverse and longitudinal parts respectively. In these expressions $t' \ll 1$ and it is the ratio of the material thickness t to the shield radius R, i.e. ($t' = t/R$), L is the length of the shield and D is called the demagnetization factor. It is the function of the L/R. The factor $\mu$ is the permeability of the magnetic material. The permeability µ is approximately 30,000 for molypermalloy and for both longitudinal and transverse fields. For two layers of shields we have

$$S_t = S_{t1} S_{t2} d' + S_{t1} + S_{t2} \qquad (4a)$$

$$S_l = 2 S_{l1} S_{l2} f d' + S_{l1} + S_{l2} \qquad (4b)$$

Where $f$ is a geometrical factor nearly constant and has value 0.75 for 2<L/R<6 and equal to about 0.9 for L/R =10. $d'$ is given by

$$d' = \frac{1}{2}[1 - (\frac{R_1}{R_2})^2] \qquad (5a)$$

or , for closely spaced shields,

$$d' \approx \Delta R/R_m \qquad (5b)$$

$\Delta R$ being the difference of the shield radius and $R_m$ the average radius of the two shields.

The transverse shielding factor of a system having more than two concentric cylinders is given by

$$S_t = \frac{1}{2}(\mu_1 \mu_2 \mu_3 ...)(t'_1 t'_2 t'_3 ...)(d'_{12} d'_{23} d'_{13} ...) \qquad (6)$$

Where the subscript $i$ characterizes the $ith$ shield and $d'_{i,i+1}$ represents the separation between shields $i$ and $i+1$. In our case the thickness of cylinder walls are $t_1 = 1.2\ mm, t_2 = 0.7\ mm, t_3 = 0.7\ mm$ respectively, thus Transverse Shielding factor for three cylinders is

$$S_t = \frac{1}{2}\mu^3(t'_1 * t'_2 * t'_3)(d'_{12} * d'_{23} * d'_{13}) \qquad (7)$$

Where $d'_{12} = \frac{1}{2}[1-\left(\frac{R_2}{R_1}\right)^2]$, $d'_{23} = \frac{1}{2}[1-\left(\frac{R_3}{R_2}\right)^2]$, $d'_{13} = \frac{1}{2}[1-\left(\frac{R_3}{R_1}\right)^2]$.

No general formula for the longitudinal shielding factor of a set of more than two cylindrical shields is available. However, for more than two cylinders an approximate expression for $S_l$ may be obtained by replacing the quantity $\mu_i$ by $4D_i\mu_i/(1+R_i/L_i)$ and the quantity $d'_{i,i+1}$ by $f_i d'_{i,i+1}$ respectively in above equation of $S_t$, where $D_i$ is magnetization factor of shield cylinder $i$. In our case we have, the Longitudinal Shielding factor for three cylinders is

$$S_l = \frac{1}{2}[(\frac{4D_1\mu 1}{1+\left(\frac{R_1}{L_1}\right)})(\frac{4D_2\mu 2}{1+\left(\frac{R_2}{L_2}\right)})(\frac{4D_3\mu 3}{1+\left(\frac{R_3}{L_3}\right)})(t'_1 t'_2 t'_3)(d'_{12} d'_{23} d'_{13})(f_1 f_2 f_3)] \qquad (8)$$

Since in our case $D_1 = D_2 = D_3 = D$ and $f_1 = f_2 = f_3 = f = 0.75$ as 2<L/R<6

Or 
$$S_l = \frac{64 S_t D^3 f^3}{\{1+\left(\frac{R_1}{L_1}\right)\}\{1+\left(\frac{R_2}{L_2}\right)\}\{1+\left(\frac{R_3}{L_3}\right)\}} \qquad (9)$$

The value of demagnetization factor D depends on the ratio L/R [2].

$$D = \frac{(1+R/L)}{2*f(1+L/R)} \qquad (10)$$

As mentioned earlier, $f$ is a geometrical factor and is nearly constant and has a value $f = 0.75$ for 2<L/R<6 and equal to about 0.9 for L/R =10. For the different values of L/R, the values of the demagnetization factor D are shown below in table-1.

Table-1

| L/R | Demagnetization factor (D) |
|---|---|
| 2 | 0.334 |
| 3 | 0.223 |
| 4 | 0.1667 |
| 5 | 0.1334 |
| 6 | 0.112 |

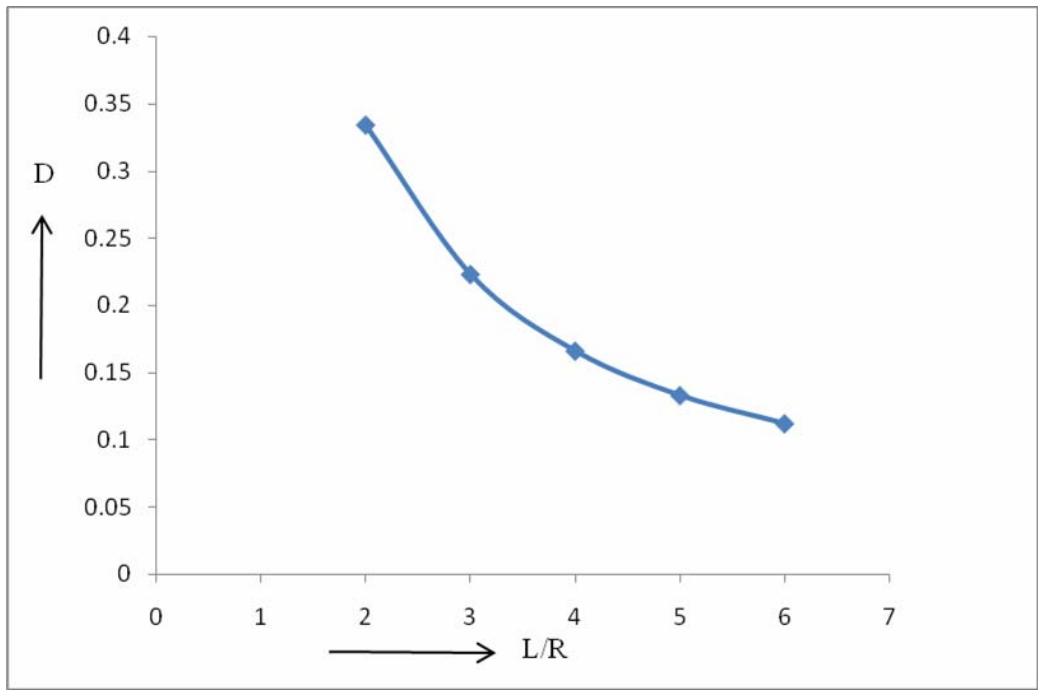

Fig. 1. Plot between diamagnetic factor D and L/R

Let us consider a typical case of the Rb physics package with three concentric layers of mu metal cylinders of lengths $L_1 = 200\ mm$, $L_2 = 190\ mm$, $L_3 = 180\ mm$ and Radii $R_1 = 65\ mm$, $R_2 = 55\ mm$, $R_3 = 45\ mm$. For the determination of the demagnetization factor D we take the average values of L and R. The average values of Length $L = \frac{L_1+L_2+L_3}{3} = 190\ mm$ and Radius $R = \frac{R_1+R_2+R_3}{3} = 55\ mm$ so from the Fig.1 the demagnetization factor D has the value 0.19323. Now the Transverse and Longitudinal shielding factors are $S_t$ = **301548.84** $S_l$ = **27375.85.** For generalizing the shielding factor for any value of length in equation (9) we vary the length $L$ =200 mm and it varies as $L = 200 - x$ and we obtain the variation in only Longitudinal shielding factor shown below in table-2

Table-2

| x(mm) | Longitudinal Sh.Factor ($S_l$) |
|---|---|
| 0 | 28274.58 |
| 10 | 27336.63 |
| 20 | 26342.64 |
| 30 | 25288.1 |
| 40 | 24168.15 |
| 50 | 22977.62 |
| 60 | 21711.12 |
| 70 | 20363.1 |
| 80 | 18928.15 |
| 90 | 17401.34 |

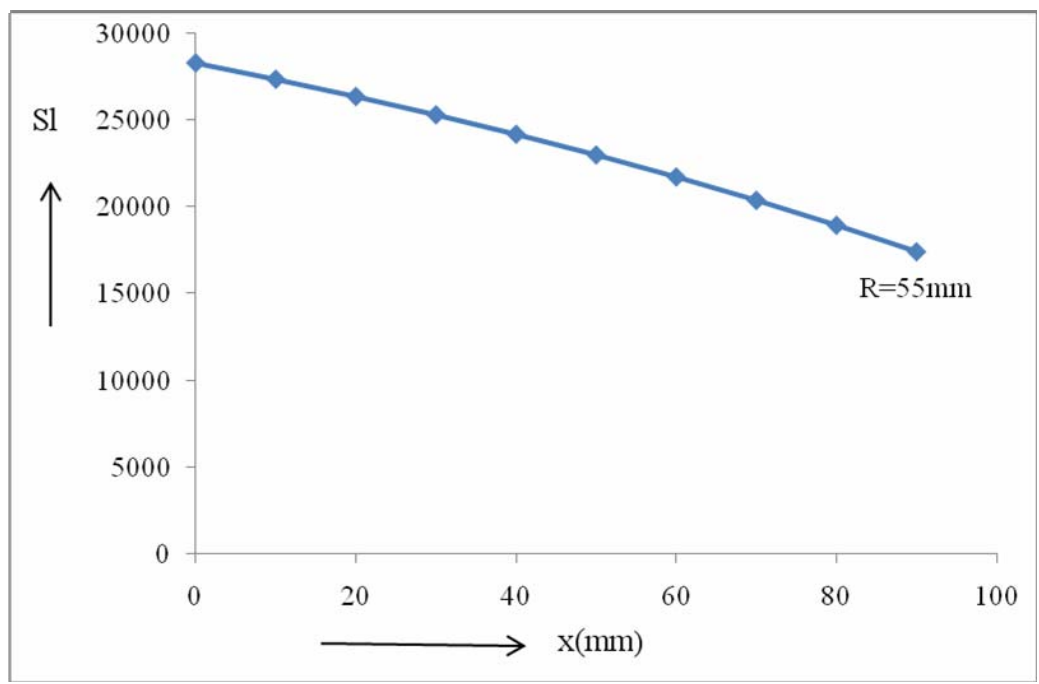

Fig. 2. **Plot between the length and Sl**

For different values of radius we also obtained the variation in the Longitudinal as well as Transverse shielding factor for different values of radius. In equation (7) and (9) we varies the R from 65 mm to descending order up to 35mm and varies it as $R_1 = 65 - y, R_2 = R_1 - 5, R_3 = R_1 - 10$ .Thus we obtained the variation in both Longitudinal and Transverse shielding factor shown below in table-3

Table-3

| y(mm) | St | $S_l$ |
|---|---|---|
| 0 | 31043 | 3434 |
| 5 | 50936 | 4612 |
| 10 | 87405 | 6325 |
| 15 | 158270 | 8894 |
| 20 | 306070 | 12886 |
| 25 | 642700 | 19363 |
| 30 | 1500875 | 30462 |

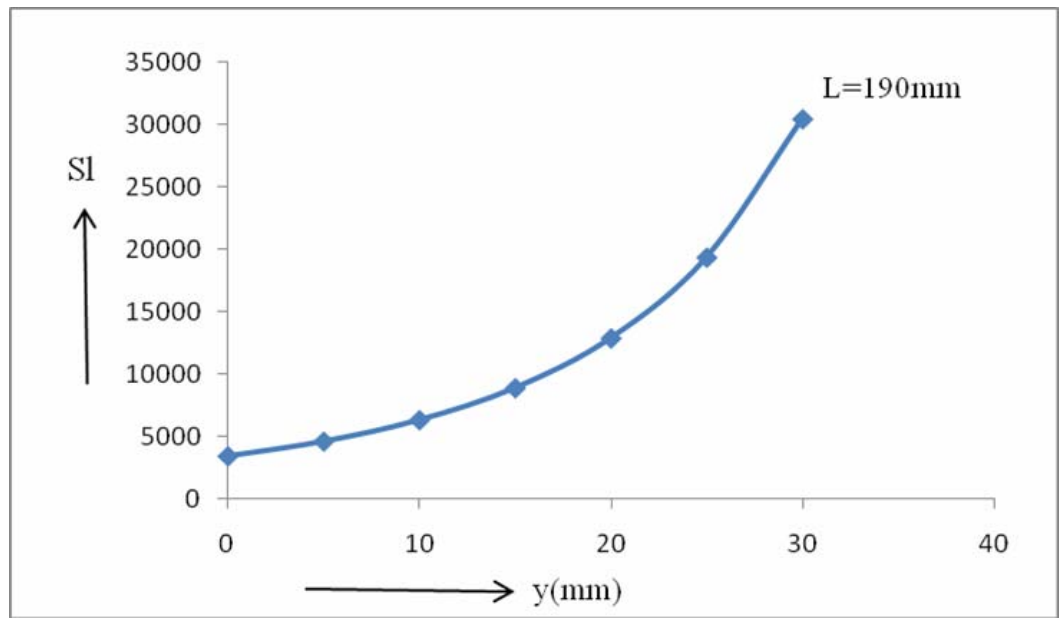

**Fig. 3. Plot between radius and longitudinal shielding factor**

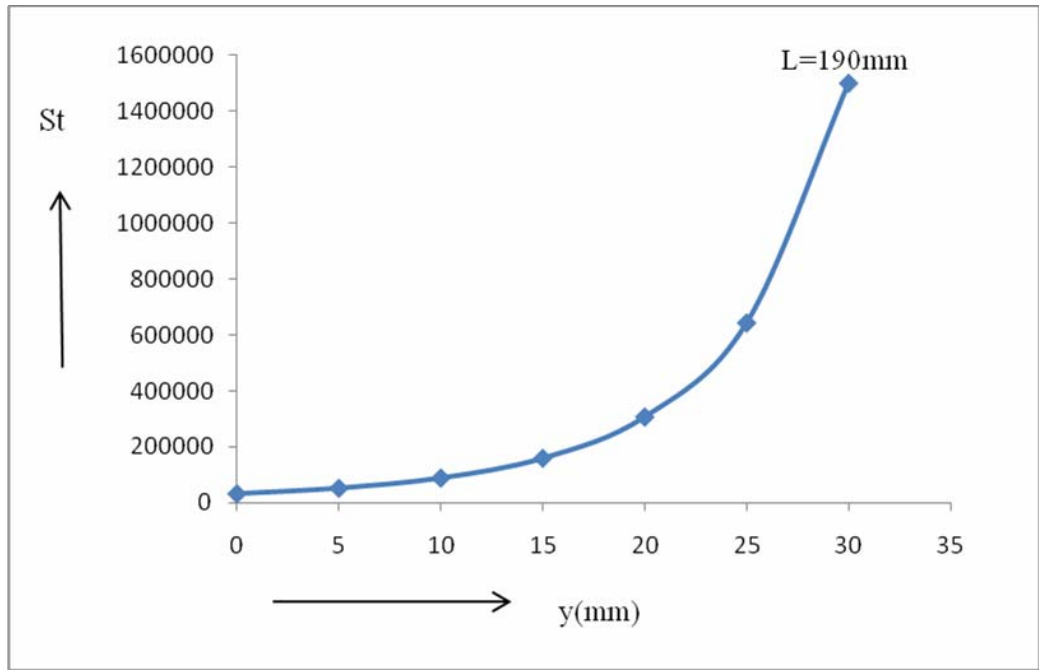

**Fig 4. Plot between radius and transverse shielding factor**

Now for small separation between shielding cylinders let consider

$R_1 = R_2 = R_3 = R$

$L_1 = L_2 = L_3 = L$

Where R and L are the average radius and average length of cylinders. Then

$$S_l = \frac{64 S_0 f^3 \left[\frac{1+R/L}{2 \cdot f(1+L/R)}\right]^3}{(1+R/L)^3} \quad (11)$$

Or

$$S_l = \frac{8 S_t}{(1+L/R)^3} \quad (12)$$

We also obtained the variation of longitudinal shielding factor with ratio L/R for 2<L/R<6 at value of $S_t = 301548.84$ as shown below in table-4

Table 4

| L/R | $S_l$ |
|---|---|
| 2 | 89347.8 |
| 3 | 37693.6 |
| 4 | 19299.1 |
| 5 | 11169.4 |
| 6 | 7033.2 |

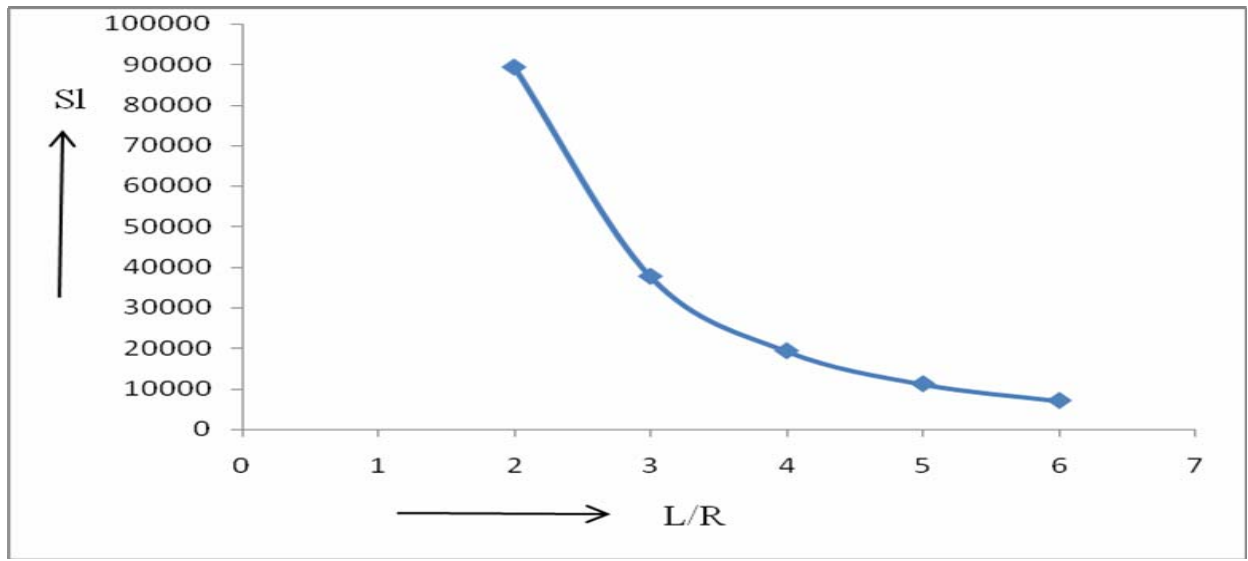

Fig. 5. Plot between $S_l$ and length to radius ratio (L/R)

Further from equation (9) we obtained variation in longitudinal shielding factor with permeability for different L/R ratio

Table 5

| μ | $S_l$ | | | | |
|---|---|---|---|---|---|
| | L/R=2 | L/R=3 | L/R=4 | L/R=5 | L/R=6 |
| 10000 | 3309 | 1396 | 715 | 413 | 260 |
| 15000 | 11167.8 | 4711 | 2412 | 1396 | 879 |
| 20000 | 26472 | 11168 | 5718 | 3309 | 2084 |
| 25000 | 51703 | 21812 | 11168 | 6463 | 4070 |
| 30000 | 89343 | 37692 | 19299 | 11168 | 7033 |
| 35000 | 141873 | 59853 | 30647 | 17735 | 11168 |
| 40000 | 211776 | 89344 | 45747 | 26473 | 16672 |

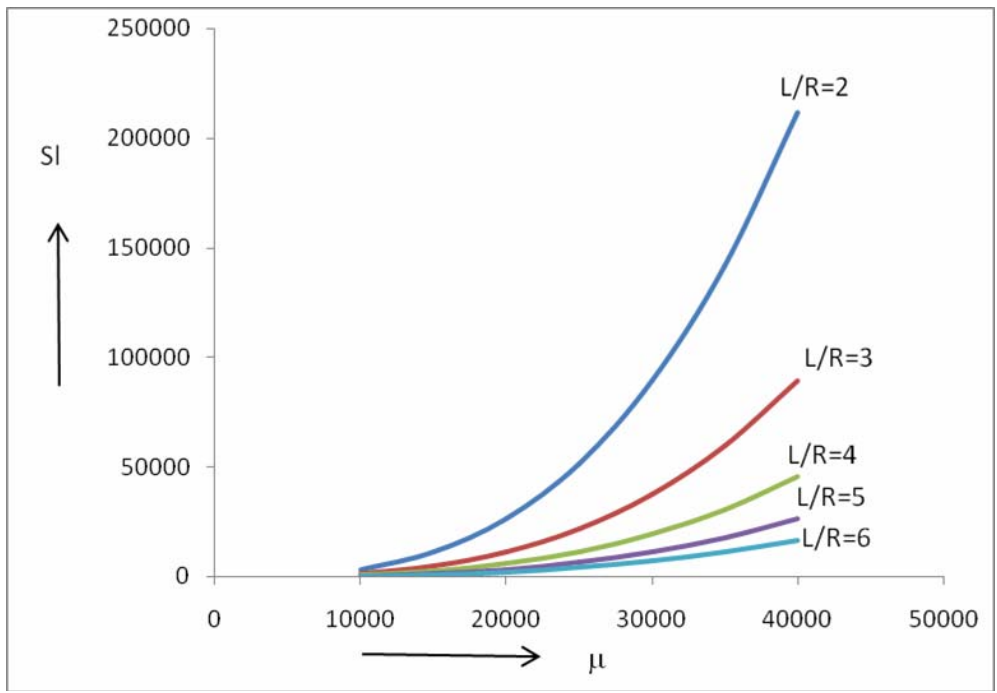

Fig. 6. Plot between longitudinal Shielding factor Sl and Permeability μ

**Programming flowchart for calculating the Transverse and Longitudinal shielding factor of the cylindrical shielding layers For Rb Atomic Clock Physics Package**

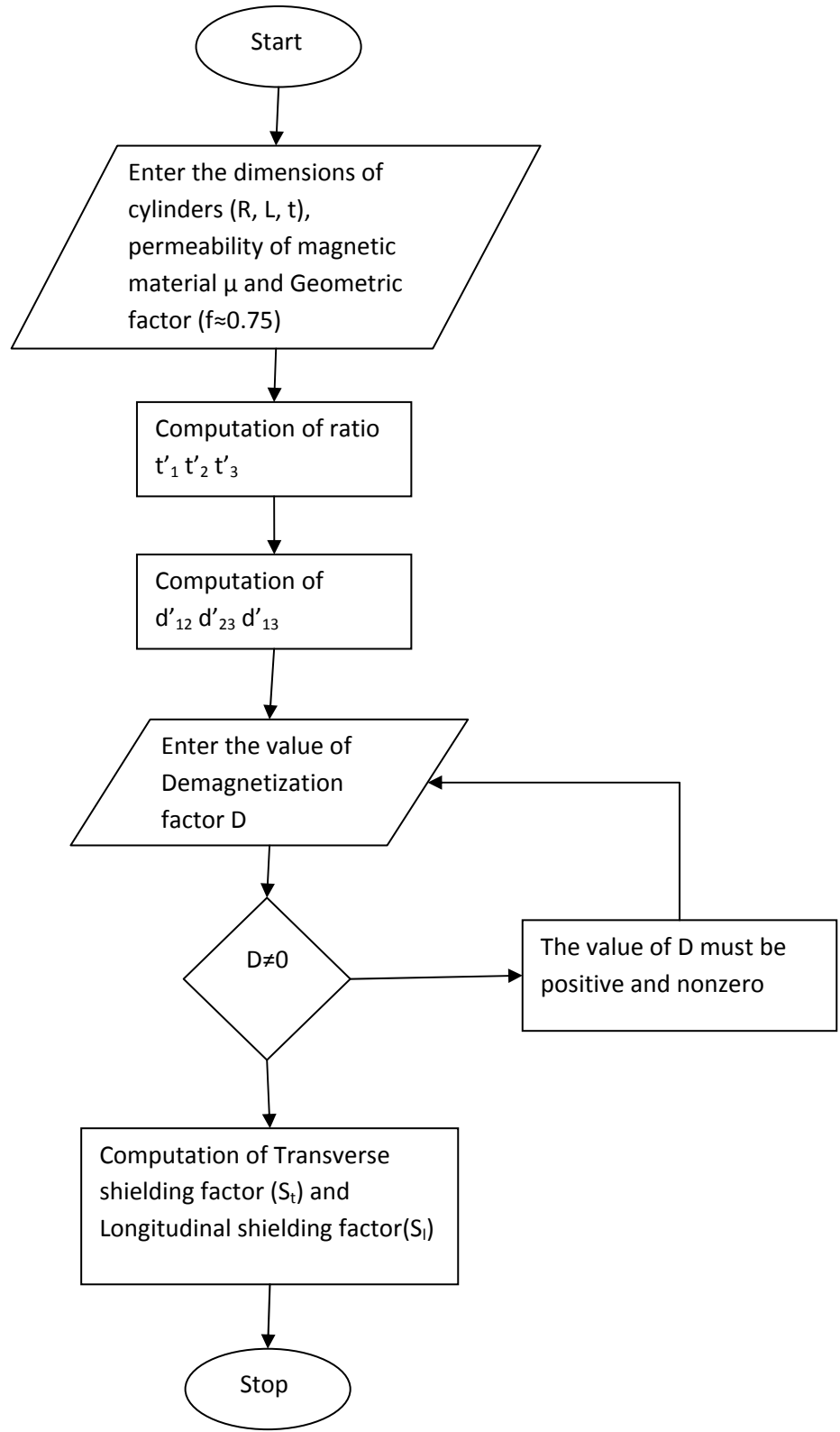

**Conclusion:** In this paper we have discussed the longitudinal and transverse magnetic shielding factors of three concentric cylindrical layers of high permeability mu metal. We have reported calculations of magnetic shielding factor (MSF) for various values of magnetic permeability and length to radius ratios and for other general cases. We observe that the MSF decreases with reducing length of the cylinders for given cross-section (fig-2).On the other hand for a given length, the MSF decreases with the increase in the cylindrical cross section (fig-3). The results reported in the paper are useful in designing magnetic shields for atomic clocks particularly Rb atomic clocks. As the ground state hyperfine levels of Rb atoms are very sensitive to the magnetic field fluctuations, the stability of the magnetic field is one of the important factor which determines the frequency stability of the Rb atomic clock.It is estimated that for obtaining frequency stability of the order of $10^{-13}$ a magnetic shielding factor of 80 dB or more is required. The calculations reported in this paper may be used for Cs atomic clocks and other application requiring magnetic shielding.